\documentclass[%
 reprint, 
 amsmath,amssymb,
 aps, physrev,
]{revtex4-2}

\usepackage{graphicx}
\usepackage{dcolumn}
\usepackage{bm}
\usepackage{amsmath}
\usepackage{orcidlink}

\usepackage{xcolor}
\usepackage{hyperref}

\usepackage{placeins}

\newcommand{\RNum}[1]{\uppercase\expandafter{\romannumeral #1\relax}}

\begin{document}


\title{\textbf{Data-Efficient Adaptation of DPA-4 Force Fields to DFT+U Energetics: A Case Study in NiO}
}%

\author{Fengyu Xie\orcidlink{0000-0002-1169-1690}}
\affiliation{
 College of Artificial Intelligence and Data Science, Suzhou Institute for Advanced Research, University of Science and Technology of China, Suzhou, Jiangsu 215123, China
}%

\author{Peiheng Jiang\orcidlink{0000-0001-9629-803X}}
\email{jiangph@xjtu.edu.cn}
\affiliation{
School of Physics, MOE Key Laboratory for Nonequilibrium Synthesis and Modulation of Condensed Matter, Xi’an Jiaotong University, Xi’an, Shaanxi 710049, China
}

\author{Zhicheng Zhong\orcidlink{0000-0003-1507-4814}}
\affiliation{
 College of Artificial Intelligence and Data Science, Suzhou Institute for Advanced Research, University of Science and Technology of China, Suzhou, Jiangsu 215123, China
}
\affiliation{
 Suzhou Lab, Suzhou, Jiangsu 215123, China
}

\date{\today}

\begin{abstract}
Foundation machine-learned force fields (MLFFs) are often pretrained on broad materials datasets whose electronic-structure conventions may not reproduce the phase energetics required for a specific correlated material. Using NiO as a case study, we examine whether incorrect source-level phase energetics can be corrected efficiently through target-level fine-tuning. Along a common structural interpolation, non-spin-polarized PBE and ferromagnetic PBE+$U$ predict opposite energetic orderings of the octahedral \textit{Oct} and square-planar \textit{Sqr} phases. Pretrained DPA-4 models adapt rapidly to the NiO PBE+$U$ surface, reaching energy and force root-mean-square errors (RMSEs) of approximately $0.5$~meV/atom and $30$~meV/$\mathrm{\AA}$, respectively, with approximately 170 PBE+$U$ labels. Crucially, models previously fine-tuned to the opposing no-$U$ surface recover the qualitative PBE+$U$ phase ordering with nearly the same target-data efficiency as models fine-tuned directly from their respective pretrained initializations. Our results show that incorrect source-level phase energetics can be reversed through target-level fine-tuning, and suggest a practical multi-fidelity strategy in which pretraining prioritizes broad, consistent, and affordable data, while compact target-level datasets impose energetics through application-specific fine-tuning.
\end{abstract}

\maketitle


\section{Introduction}

In transition-metal compounds, conventional semilocal density-functional theory (DFT) may over-delocalize the $d$ electrons and inadequately describe the associated magnetic moments. Corrective approaches such as DFT+$U$ are therefore frequently employed to improve the description of localization, magnetism, and relative phase energetics\cite{anisimovBandTheoryMott1991, cococcioniLinearResponseApproach2005,jainFormationEnthalpiesMixing2011}. In some cases, these corrections are not merely quantitative. The relative stability of competing crystal structures can be sensitive to the treatment of localized electronic states, and the inclusion or choice of an on-site interaction can even reverse their energetic ordering\cite{spagnoliDensityFunctionalTheory2010,longEvaluatingOptimal32020,kamInterplayElectronLocalization2025}. A machine-learned force field (MLFF) trained on uncorrected DFT calculations will generally inherit the corresponding reference-level errors and may consequently predict incorrect phase stability.

Nevertheless, using DFT+$U$ for large-scale MLFF pretraining may introduce a separate difficulty. Commonly used materials databases, such as Materials Project (MP)\cite{jainCommentaryMaterialsProject2013,hortonAcceleratedDatadrivenMaterials2025}, Alexandria\cite{schmidtDataset175kStable2022,schmidtMachineLearningAssistedDeterminationGlobal2023} and OMat24\cite{barros-luqueOpenMaterials20242026a} may combine calculations performed under composition-dependent DFT and DFT+$U$ conventions\cite{zhouFirstprinciplesPredictionRedox2004, jainFormationEnthalpiesMixing2011}. Although these conventions were often designed to improve agreement between calculated and experimental thermodynamics for transition-metal compounds, they effectively mix labels generated by different Hamiltonians into a single training set. Consequently, an MLFF must reconcile potentially incompatible energetic descriptions across chemical environments, which may lead to nonphysical interpolation. Recent work\cite{warfordBetterImpactSelective2026a} has investigated the effect of such dataset inconsistency in depth and argued that consistently generated data without $U$ could be more suitable for pretraining foundation MLFFs than datasets containing heterogeneous $U$ conventions.

This proposal naturally suggests a separation of roles between pretraining and system-specific fine-tuning. During pretraining, an MLFF learns broadly transferable chemical interactions from an affordable and consistently parameterized reference DFT dataset. Quantitative energetics for a particular correlated material can subsequently be imposed through fine-tuning with a compact dataset at a more appropriate theory level, such as DFT+$U$ or hybrid-functional DFT. For this strategy to be practically useful, an MLFF pretrained without $U$ must remain adaptable even when its pretraining encodes qualitatively incorrect phase energetics. Our previous work demonstrated that foundation MLFFs can be fine-tuned for specific materials with remarkable data efficiency\cite{wangPretrainingFinetuningDistillation2025}. However, it remains unclear whether prior fine-tuning to an incorrect potential-energy surface impedes subsequent adaptation, and whether a small target dataset can efficiently reverse a strongly learned energetic preference.

\begin{figure}[htbp]
    \centering
    \includegraphics[width=1.0\linewidth]{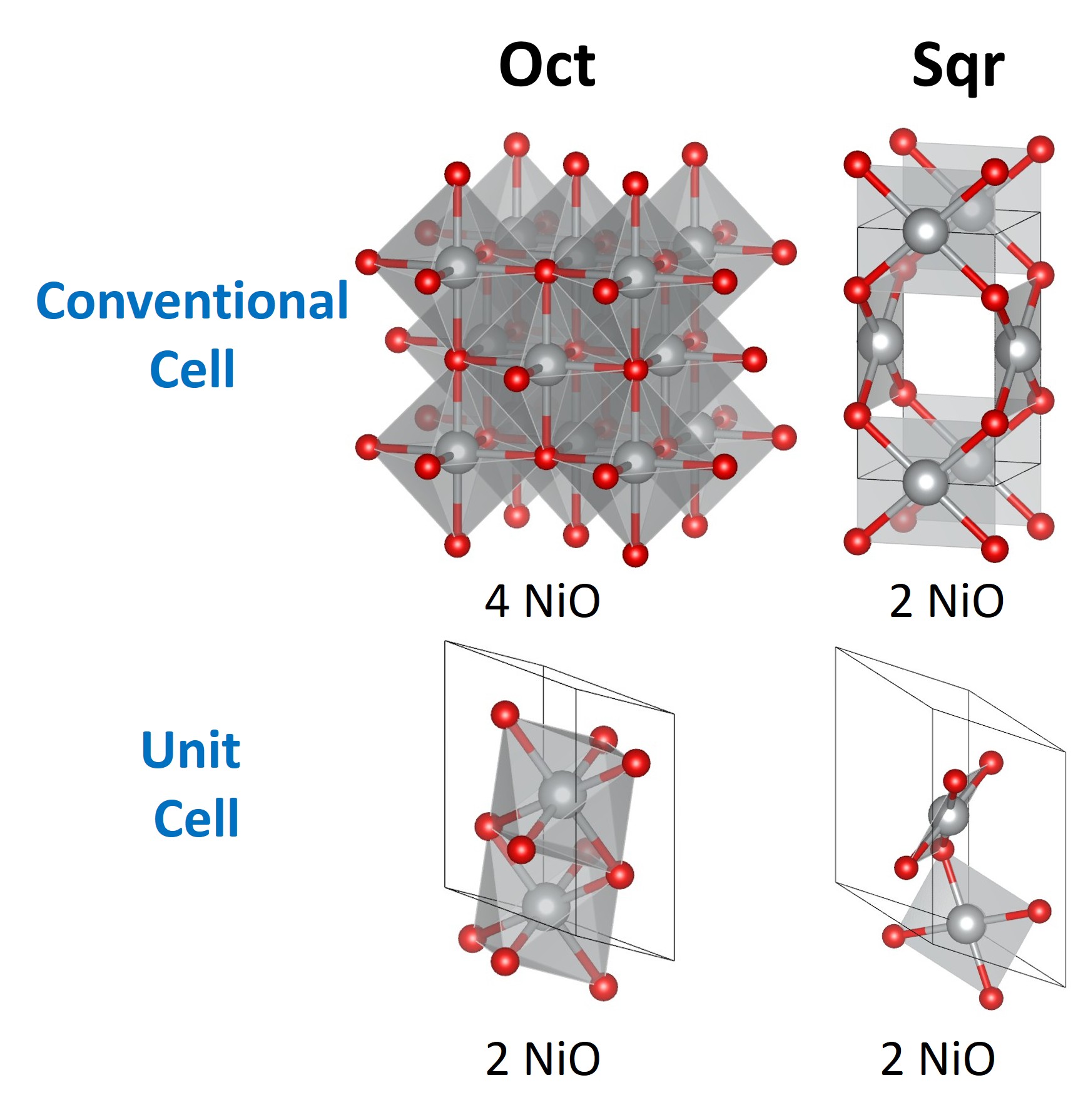}
    \caption{Crystal structures of NiO with octahedral (\textit{Oct}) and square-planar (\textit{Sqr}) Ni coordination. The upper row shows the conventional rocksalt-like cubic cell of \textit{Oct}-NiO containing four formula units (left) and the conventional PdO-like tetragonal cell of \textit{Sqr}-NiO containing two formula units (right). The lower row shows the corresponding triclinic cells of \textit{Oct}-NiO (left) and \textit{Sqr}-NiO (right), each containing two formula units. The two-formula-unit triclinic cells were used for the structural interpolation and as the parent cells for constructing the $2\times2\times2$ AIMD supercells.
    }\label{fig:phases}
\end{figure}

Here, we investigate these questions using NiO, an archetypal strongly correlated transition-metal oxide, as a case study. We consider a rocksalt-like structure with octahedrally coordinated Ni, denoted \textit{Oct}, and a PdO-like tetragonal structure with square-planar Ni coordination, denoted \textit{Sqr} (Figure~\ref{fig:phases}). Along a common structural interpolation between these phases, non-spin-polarized PBE and spin-polarized PBE+$U$ predict opposite relative energetics. The PBE+$U$ trend is qualitatively supported by HSE06 and is preserved under both ferromagnetic (FM) and antiferromagnetic (AFM) ordering. This pronounced reversal provides a stringent test of whether an MLFF carrying an incorrect source-level energetic preference can be efficiently adapted to a target-level surface.

We fine-tune several pretrained DPA models\cite{zhangGraphNeuralNetwork2026, liDPA4PushingAccuracyCost2026} using a NiO dataset generated with ferromagnetic PBE+$U$ under consistent computational settings. With 170 training configurations, the fine-tuned models reproduce the PBE+$U$ energies, forces, virials, and the \textit{Oct}--\textit{Sqr} phase ordering. For the DPA-4 models, fine-tuning reduces the energy root-mean-square error (RMSE) from approximately 15--20 meV/atom to below 0.5 meV/atom and the force RMSE from approximately 125--175 meV/$\mathrm{\AA}$ to below 30 meV/$\mathrm{\AA}$. Most importantly, DPA-4 models previously fine-tuned on non-spin-polarized PBE data, and therefore encoded the opposite \textit{Oct}--\textit{Sqr} energetic preference, can be transferred to the ferromagnetic PBE+$U$ surface with nearly the same data efficiency as models fine-tuned directly from the corresponding pretrained initialization.

Our results demonstrate that incorrect source-level phase energetics do not necessarily make a pretrained MLFF unsuitable for subsequent target-level adaptation. More broadly, they suggest a practical multi-fidelity strategy for developing and using foundation MLFFs. Pretraining can prioritize broad configurational coverage, consistency of the DFT labels, and computational affordability, whereas compact target-level datasets can provide the electronic and energetic fidelity required for a specific application. This suggestion does not imply that arbitrary source-level errors are harmless. Rather, the present NiO example shows that target-level phase ordering need not be reproduced during pretraining for the resulting MLFF representation to remain useful. Foundation MLFFs should therefore be regarded as transferable priors that require target-specific validation and, where necessary, appropriate target-level fine-tuning.

\section{Methods}
\subsection{DFT calculations}
All DFT calculations, including structural relaxations and single-point calculations, were carried out using the Vienna \textit{ab initio} Simulation Package (VASP) with the projector-augmented-wave (PAW) method\cite{kresseInitioMolecularDynamics1993,kresseEfficiencyAbinitioTotal1996,kresseEfficientIterativeSchemes1996,kresseUltrasoftPseudopotentialsProjector1999}. A plane-wave cutoff of 520 eV was used. Input files were generated using the default PBE\_52 pseudopotentials provided through pymatgen \cite{ongPythonMaterialsGenomics2013}, with Ni 3p electrons treated as valence electrons. Electronic convergence was set to $5 \times 10^{-6}$ eV/atom, and ionic steps were converged to 0.02 eV/$\mathrm{\AA}$ for all structural relaxations. For all DFT+$U$ calculations, we consistently employed the rotationally invariant form introduced by Dudarev \textit{et al.}\cite{dudarevElectronenergylossSpectraStructural1998a} with $U_{\mathrm{eff}}=U-J=6.2$~eV applied to the Ni 3d orbitals, following the convention implemented in pymatgen.

In order to compare \textit{Oct}--\textit{Sqr} energetics under different correlation treatments, we employed PBE\cite{perdewGeneralizedGradientApproximation1996,perdewGeneralizedGradientApproximation1997}, PBE+$U$, r2SCAN\cite{furnessAccurateNumericallyEfficient2020} and HSE06\cite{krukauInfluenceExchangeScreening2006}. No additional Hubbard correction was applied to r2SCAN and HSE06. In these computations, we used triclinic cells containing two NiO formula units (Figure~\ref{fig:phases}, lower panels), and applied a $\Gamma$-centered $8\times8\times5$ \textit{k}-point mesh. The \textit{Oct} and \textit{Sqr} endpoint structures were first relaxed using non-spin-polarized PBE without U. Intermediate configurations were then generated by linearly interpolating both the atomic coordinates and lattice matrices. The resulting fixed interpolation pathway was evaluated using single-point calculations at each electronic-structure level.

To obtain PBE and FM PBE+$U$-level datasets for fine-tuning MLFFs, we performed \textit{ab initio} molecular dynamics (AIMD) in the isothermal--isobaric (NPT) ensemble at $T=300, 600, 900$~K and $P=1$~bar. Both the cell volume and the cell shape were allowed to vary. The AIMD trajectories were propagated using the Langevin thermostat and barostat, with lattice friction coefficients of 10~ps$^{-1}$ and a fictitious lattice mass of 1000 atomic mass units. At each temperature, we performed two simulations, initialized from the \textit{Oct} and the \textit{Sqr} structures, respectively. Each simulation lasted 250 steps with a time step of 4.0~fs. From the resulting PBE trajectories of the \textit{Oct} and the \textit{Sqr} phases, we extracted 750 electronically converged frames for each phase. From the FM PBE+$U$ trajectories, we extracted 750 and 250 electronically converged frames for the \textit{Oct} and the \textit{Sqr} phases, respectively. These extracted frames subsequently formed the datasets for fine-tuning and validating MLFFs. For AIMD, the two-formula-unit cells were replicated $2\times2\times2$, yielding supercells containing 16 NiO formula units to sample a broader range of local configurations.

\subsection{Dataset preparation}
For the datasets generated from AIMD calculations, we first held out 10\% of all configurations at random for testing. Training configurations were then selected from the remaining configurations by maximizing the information entropy implemented in QUESTS\cite{QUESTS20240621Documentation}. The information entropy of a set of $n$ structures ($H$) is defined as\cite{schwalbe-kodaModelfreeEstimationCompleteness2025}
\begin{equation}
    H(\mathbf{\{X\}}) = - \frac{1}{n} \sum_{i=1}^{n} \log \left[\frac{1}{n} \sum_{j=1}^{n} K_h(\mathbf{X}_i, \mathbf{X}_j) \right],
\end{equation}
where $\mathbf{X}_i$ denotes the vectorized local atomic features of structure $i$, $K_h$ is a Gaussian kernel, and $h$ is the kernel bandwidth, set to $h=0.015$ in this work. The local atomic features were constructed using 32 channels and a radial cutoff of 5.0~$\mathrm{\AA}$. The differential entropy gain upon adding a candidate structure $\mathbf{Y}$ into the existing dataset $\mathbf{\{X\}}$ is given by
\begin{equation}
    \delta H(\mathbf{Y}|\mathbf{\{X\}}) = -\log \left[\sum_{i=1}^{n} K_h(\mathbf{Y}, \mathbf{X}_i)\right].
\end{equation}
All DFT-labeled structures were ranked by $\delta H$, and the top $N_{\mathrm{train}}$ configurations were selected to maximize the information diversity of the training set for a given training-set size. 

Using this approach, we selected $N_\mathrm{train}=660$ configurations from the PBE dataset to fit the pretrained MLFFs accurately to the non-spin-polarized PBE surface. For the PBE+$U$ dataset, we varied $N_\mathrm{train}$ from 50 to 290 in increments of 40 to investigate how rapidly test errors decrease with increasing training set size. For each $N_\mathrm{train}$, three independent train--test splits were generated in order to assess the reproducibility of test errors across different data splits.

\subsection{Foundation MLFFs and fine-tuning}\label{sec:mlff}
We compared four foundation MLFFs: the \textit{MPtraj} branch and the \textit{OMat24} branch of DPA-3.1-3M\cite{zhangGraphNeuralNetwork2026,Models343DPA313M}, denoted as DPA3-MP and DPA3-OMAT, respectively, and two DPA4\cite{liDPA4PushingAccuracyCost2026} models, denoted as DPA4-Air-MP and DPA4-Neo-OMAT. The neighbor cutoff radii were set to 6~$\mathrm{\AA}$ for all these models. Models were fine-tuned and evaluated using deepmd-kit\cite{zengDeePMDkitV3MultipleBackend2025}.

Models were trained for approximately 120 epochs, corresponding to $\sim120\times N_{\mathrm{train}}/N_\mathrm{batch}$ steps. Automatic batch sizing was used with a target of 128 atoms per batch. The gradient norm was clipped to a maximum of 1.0. For DPA-3, the learning rate decayed exponentially from 0.001 to $3\times10^{-5}$ over the full training schedule at an interval of $N_\mathrm{train}$ steps.  The DPA-3 loss function combined energy, force, and virial mean square error (MSE) terms. Their weights were varied from 0.2:100:0.02 at the beginning of training to 20:60:1 at the end. For DPA-4, the learning rate followed a warmup-stable-decay (WSD) schedule, starting at $1.4\times10^{-4}$, warming up to $7.0\times10^{-4}$ in $N_\mathrm{train}$ steps, remaining stable until 65\% of the total training steps had been completed, after which it decayed cosinusoidally to $1\times10^{-6}$. The DPA-4 loss combined energy, force, and virial mean absolute error (MAE) terms with fixed weights of 1:1:0.25, respectively.

All DFT data and fine-tuned MLFFs are available on Figshare at \url{https://doi.org/10.6084/m9.figshare.33191403}.

\begin{figure}[htbp]
    \centering
    \includegraphics[width=1.0\linewidth]{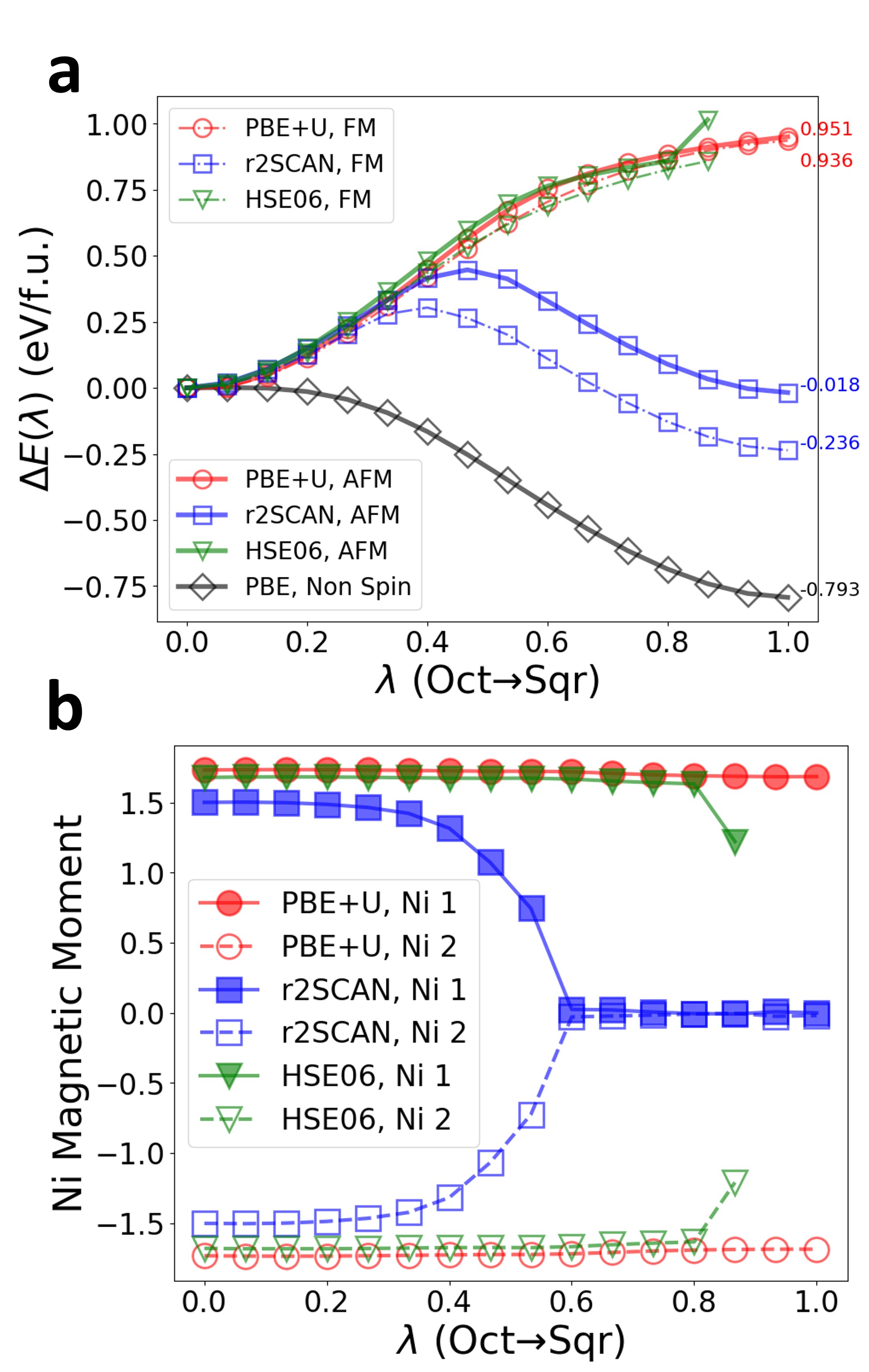}
    \caption{Correlation-sensitive energetics and site-resolved magnetic moments along the \textit{Oct}--\textit{Sqr} interpolation path in NiO. (a) Energy per NiO formula unit relative to the \textit{Oct} configuration ($\Delta E(\lambda)=E(\lambda)-E(0)$) as a function of the interpolation coordinate $\lambda$ between the PBE-relaxed \textit{Oct} ($\lambda=0$) and \textit{Sqr} ($\lambda=1$) endpoints. The configurational energies were evaluated on the same set of fixed configurations. PBE+$U$, r2SCAN, and HSE06 results are shown as red circles, blue squares, and green triangles, while solid and dash-dotted lines denote FM and AFM ordering, respectively. Non-spin-polarized PBE results are shown as open black diamonds connected by a solid line. \textit{Sqr}-endpoint values, $\Delta E(\lambda=1)$, are annotated for the consistently converged PBE, PBE+$U$ and r2SCAN calculations. The final two HSE06 points are omitted because magnetic solutions could not be converged consistently. (b) Local magnetic moments on the two Ni sites from the AFM calculations as functions of $\lambda$. Colors and marker shapes follow panel (a). Ni 1 and Ni 2 are distinguished by filled and open markers and by solid and dashed lines, respectively.
    }\label{fig:phase_energetics}
\end{figure}

\section{Results} 
\subsection{Correlation-sensitive phase energetics in NiO}

Figure~\ref{fig:phase_energetics} compares the relative energies of NiO configurations along a common structural interpolation from \textit{Oct} to \textit{Sqr}. We first relaxed the endpoint structures using non-spin-polarized PBE (black diamonds, solid line), and generated the intermediate configurations by linear interpolation. We subsequently performed single-point calculations with PBE+$U$, r2SCAN, and HSE06 under FM and AFM orderings on the same set of fixed geometries, in order to isolate differences arising from the electronic-structure method from those caused by method-dependent structural relaxation. The interpolation therefore represents a fixed configurational path rather than a minimum-energy path. The final two HSE06 configurations are omitted because consistent magnetic solutions could not be reached.

The calculated energy profiles depend qualitatively on the treatment of electronic correlation. At the fixed PBE-relaxed endpoint geometries, non-spin-polarized PBE places \textit{Sqr} 0.793~eV per NiO formula unit (eV/f.u.) below \textit{Oct}. PBE+$U$ reverses this ordering, placing \textit{Sqr} 0.936 and 0.951 eV/f.u.\ above \textit{Oct} under FM and AFM orderings, respectively. HSE06 exhibits the same destabilization toward \textit{Sqr} as PBE+$U$ over the portion of the path for which consistent magnetic solutions were obtained. The qualitative agreement between PBE+$U$ and the independent, higher-cost HSE06 reference supports the use of PBE+$U$ as the target electronic-structure level in the subsequent transfer experiments. 

Even though pretraining exclusively on r2SCAN data has recently been suggested as a possible route toward improving the thermodynamic accuracy and zero-shot phase predictions of foundation MLFFs\cite{kaplanFoundationalPotentialEnergy2025,kunerMPALOER2SCANDataset2025a,shinagawaMatlantisPFPV8Universal2026}, r2SCAN pretraining cannot by itself guarantee the elimination of material-dependent errors in the underlying reference theory. In the present case, r2SCAN without $U$ places \textit{Sqr} below \textit{Oct} by $0.018$~eV/f.u.\ under FM ordering and by $0.236$~eV/f.u.\ under AFM ordering along the \textit{Oct}--\textit{Sqr} interpolation, in qualitative disagreement with both PBE+$U$ and HSE06. As shown in Fig.~\ref{fig:phase_energetics}(b), the change in the r2SCAN energy profile coincides with a collapse of the local Ni moments, indicating that r2SCAN does not preserve the localized-moment branch throughout the interpolation. Thus, although r2SCAN may improve pretraining fidelity on average, r2SCAN pretraining should not be assumed to provide universally reliable zero-shot phase energetics. System-specific validation and fine-tuning at the target reference level remain essential.

The qualitative reversal makes NiO a stringent test of adaptation from an incorrect source-level energetic preference to the target PBE+$U$ surface. Because the differences between the FM and AFM PBE+$U$ profiles are much smaller than the differences among electronic-structure methods and both yield the same qualitative \textit{Oct}--\textit{Sqr} ordering, we use the consistently converged FM PBE+$U$ branch as a single, well-defined target surface for the subsequent MLFF transfer experiments.

\subsection{Data-efficient fine-tuning of pretrained MLFFs to PBE+$U$}

\begin{figure}[htbp]
    \centering
    \includegraphics[width=1.0\linewidth]{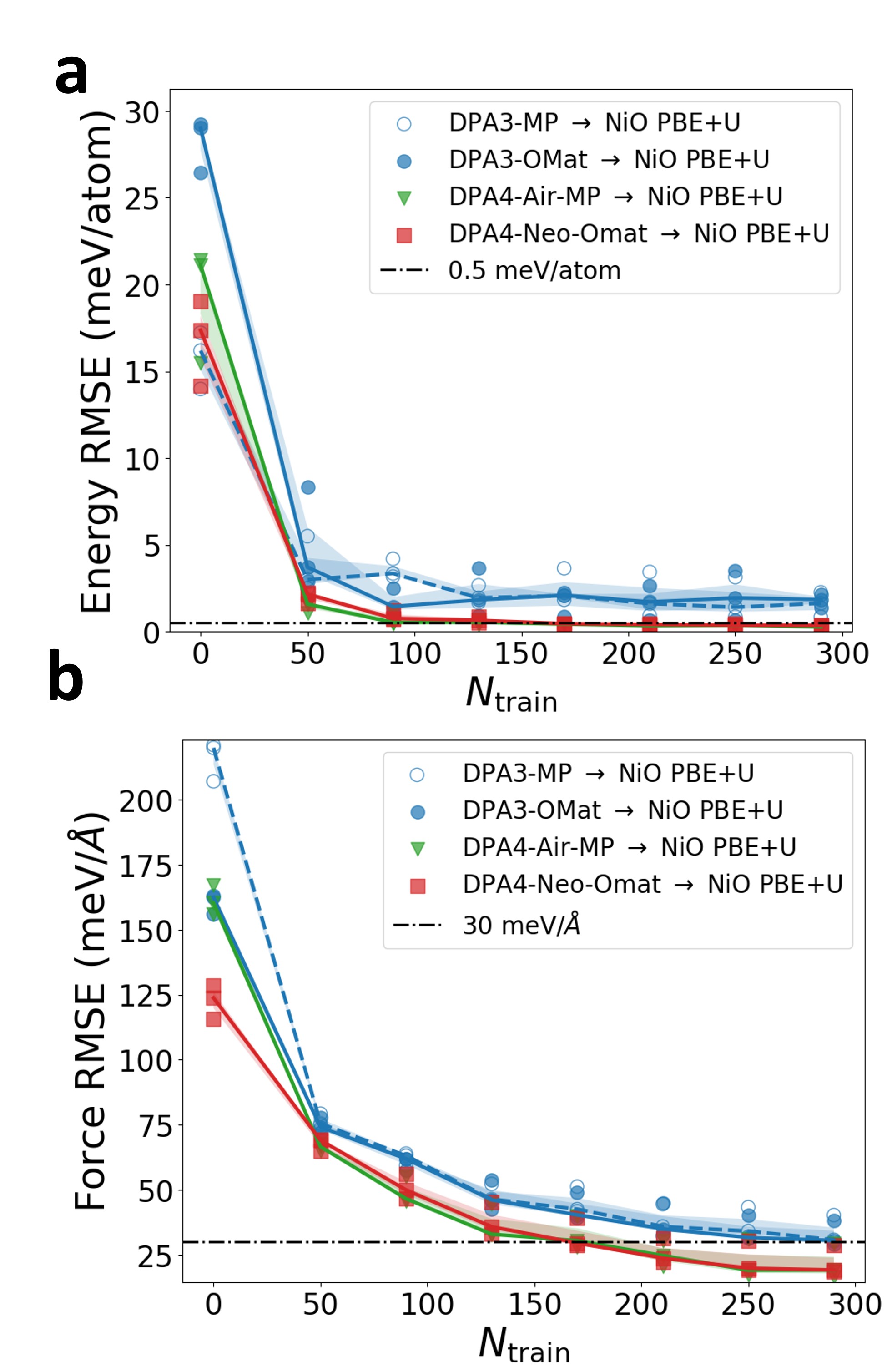}
    \caption{Data-efficient adaptation of pretrained MLFFs to the NiO PBE+$U$
        reference surface. Test-set RMSEs for (a) energies and (b) forces as functions of the number of PBE+$U$ training configurations, $N_{\mathrm{train}}$. Results at $N_{\mathrm{train}}=0$ correspond to zero-shot predictions of the pretrained models. Symbols show results from individual train--test splits. Open blue circles along with dashed lines denote DPA3-MP, while the filled blue circles, green triangles and red squares along with solid lines denote DPA3-OMAT, DPA4-Air-MP and DPA4-Neo-OMAT, respectively. Shaded regions indicate the interquartile range across independent train--test splits at each $N_{\mathrm{train}}$. The horizontal dash-dotted lines indicate energy and force RMSEs of 0.5 meV/atom in panel (a) and 30 meV/$\mathrm{\AA}$ in panel (b).
        }\label{fig:pretrain_transfer}
\end{figure}

We next examine whether pretrained MLFFs can be efficiently adapted to the FM PBE+$U$ reference surface. Figure~\ref{fig:pretrain_transfer} shows the energy and force RMSEs on the held-out PBE+$U$ test set as functions of the number of target-level training configurations, $N_{\mathrm{train}}$. The results at $N_{\mathrm{train}}=0$ correspond to zero-shot predictions. We compare four pretrained models spanning three architectures: DPA3, DPA4-Air, and DPA4-Neo. For DPA3, we consider two pretrained branches, DPA3-MP and DPA3-OMAT, trained on the MP\cite{jainCommentaryMaterialsProject2013} and OMat24\cite{barros-luqueOpenMaterials20242026a} datasets, respectively. For DPA4-Air and DPA4-Neo, we use the separately released DPA4-Air-MP and DPA4-Neo-OMAT checkpoints, pretrained on MP and OMat24, respectively. For each $N_{\mathrm{train}} > 0$, we train and evaluate each model using three independent train--test splits; for the zero-shot case we evaluate the pretrained models on the corresponding test sets.

All four models improve substantially upon fine-tuning with a small amount of PBE+$U$ data. In particular, the zero-shot energy RMSEs of the DPA-4 models are approximately  15--20 meV/atom and decrease to approximately 1--2 meV/atom with only 50 training configurations and approach 0.5 meV/atom after fine-tuning on 170 configurations. Their force RMSEs similarly decrease from approximately 125--175 meV/$\mathrm{\AA}$ for zero-shot predictions to approximately 30 meV/$\mathrm{\AA}$ with 170 training configurations and improve further with additional training data. These results show that foundation MLFFs pretrained on large materials datasets can be transferred efficiently to a consistently defined, system-specific PBE+$U$ surface, despite differences between the electronic-structure conventions represented during pretraining and the consistently defined target level used for fine-tuning.

Under the present fine-tuning protocol (see Methods Section~\ref{sec:mlff}), the tested DPA-4 models exhibit greater data efficiency than the tested DPA-3 models. Although the DPA-3 models also benefit from target-level fine-tuning, their energy RMSEs remain approximately 1--2~meV/atom over much of the investigated range, and their force RMSEs reach approximately 30 meV/$\mathrm{\AA}$ only for the largest training subsets ($N_{\mathrm{train}}=290$). In comparison, both DPA-4 models reach lower energy and force RMSEs with fewer PBE+$U$ labels.

\subsection{Recovery after no-$U$ fine-tuning}

\begin{figure*}[htbp]
    \centering
    \includegraphics[width=1.0\linewidth]{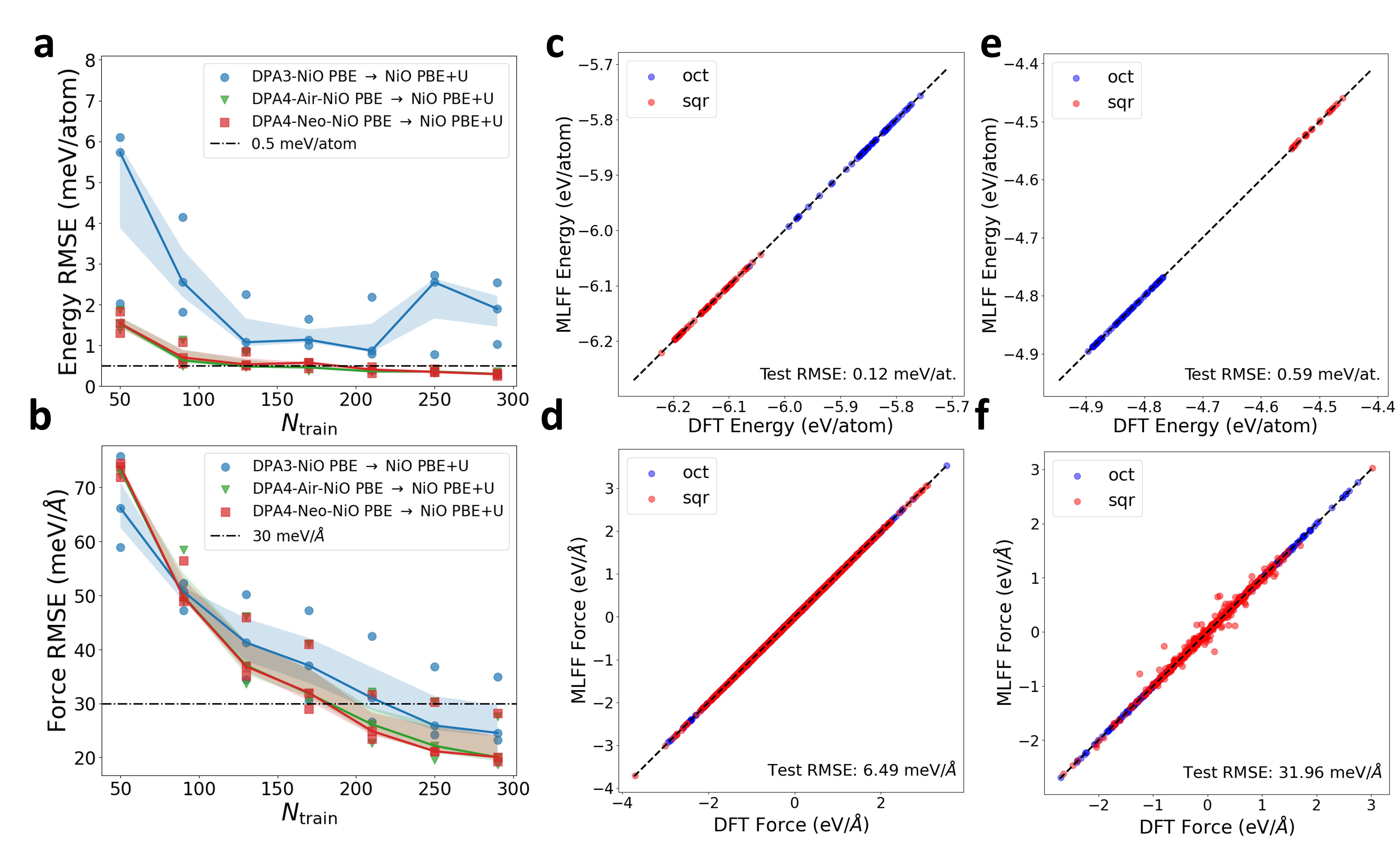}
    \caption{Second-stage adaptation from the non-spin-polarized PBE surface without $U$ to the FM PBE+$U$ target surface. (a--b) Test-set RMSEs for (a) energies and (b) forces as functions of the number of FM PBE+$U$ training configurations, $N_\mathrm{train}$, used in the second fine-tuning stage. Before this stage, each model was fine-tuned on non-spin-polarized NiO PBE data without $U$. Symbols show results from independent train--test splits. Blue circles, green triangles and red squares denote DPA3, DPA4-Air and DPA4-Neo, respectively; solid lines show the corresponding trends. Shaded regions indicate the interquartile range across independent train--test splits at each $N_{\mathrm{train}}$. The horizontal dash-dotted lines indicate energy and force RMSEs of 0.5 meV/atom in panel (a) and 30 meV/$\mathrm{\AA}$ in panel (b). (c--f) Parity plots of MLFF predictions against the corresponding DFT labels. The dashed diagonal lines denote ideal agreement. (c) Energy and (d) force parity plots for DPA4-Neo after the first-stage fine-tuning on 660 non-spin-polarized PBE configurations without U. (e) Energy and (f) force parity plots for DPA4-Neo after second-stage fine-tuning on 170 FM PBE+$U$ configurations. 
    }\label{fig:recovery}
\end{figure*}

We further test a more demanding transfer scenario in which the pretrained models are first fine-tuned on non-spin-polarized PBE data without $U$ and subsequently transferred to the FM PBE+$U$ target surface. The intermediate fine-tuning stage deliberately fits the models to the potential-energy surface that favors \textit{Sqr} over \textit{Oct}, allowing us to test whether prior learning of a qualitatively incorrect phase preference impedes subsequent target-level adaptation.

To verify accurate fitting to the no-$U$ reference surface, we examine DPA4-Neo-OMAT, which was first fine-tuned using 660 non-spin-polarized PBE configurations. As shown in Figs.~\ref{fig:recovery}(c) and \ref{fig:recovery}(d), the resulting model reproduces the test-set no-$U$ energies and forces with RMSEs of 0.12 meV/atom and 6.49 meV/$\mathrm{\AA}$, respectively. The intermediate model therefore reproduces the no-$U$ surface accurately, rather than merely acquiring only a weak source-level bias.

Figures~\ref{fig:recovery}(a) and \ref{fig:recovery}(b) show the test errors obtained when the no-$U$-adapted models are further fine-tuned using FM PBE+$U$ data. Despite their prior alignment with the opposing energetic preference, both DPA-4 models adapt rapidly to the PBE+$U$ surface. Their RMSEs decrease with increasing training set size at rates comparable to those obtained by direct fine-tuning from the corresponding foundation models in Fig.~\ref{fig:pretrain_transfer}. With 170 PBE+$U$ training configurations, DPA4-Neo reaches energy and force RMSEs of 0.59 meV/atom and 31.96 meV/$\mathrm{\AA}$, respectively, as further shown in Figs.~\ref{fig:recovery}(e) and \ref{fig:recovery}(f). Therefore, prior alignment with the no-$U$ surface does not substantially increase the number of target-level configurations required by DPA-4.

\begin{figure}[htbp]
    \centering
    \includegraphics[width=1.0\linewidth]{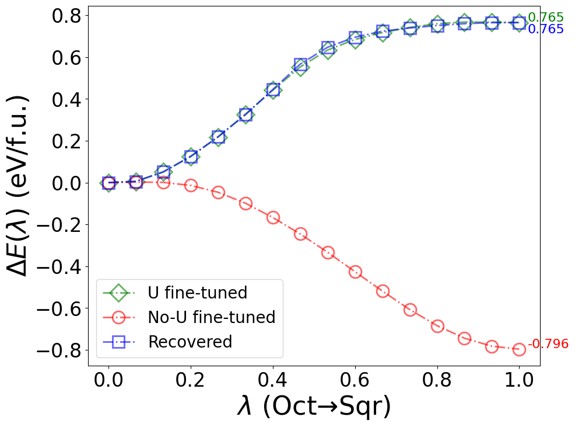}
    \caption{Predicted relative energies, $\Delta E(\lambda)$, along the common \textit{Oct}--\textit{Sqr} structural interpolation path ($\lambda$), where $\lambda=0$ and 1 correspond to \textit{Oct} and \textit{Sqr}, respectively. For each model, energies are referenced to its prediction for \textit{Oct} at $\lambda=0$, such that $\Delta E(\lambda)=E(\lambda)-E(0)$. Red circles denote predictions from the model fine-tuned on non-spin-polarized PBE data without U, while green diamonds and blue squares denote the model fine-tuned directly from its pretrained initialization to PBE+$U$ and the model first fine-tuned on PBE without $U$ and subsequently fine-tuned on PBE+$U$, respectively. All predictions are obtained with DPA4-Neo. 
    }\label{fig:phase_recovery}
\end{figure}

More importantly, second-stage fine-tuning corrects not only the aggregate test errors but also the qualitative phase energetics. Figure~\ref{fig:phase_recovery} compares the relative-energy profiles predicted by the no-$U$-adapted, directly PBE+$U$-fine-tuned, and two-stage-fine-tuned DPA4-Neo models along the common \textit{Oct}--\textit{Sqr} interpolation path. The no-$U$-adapted model places the \textit{Sqr} endpoint $0.796$~eV/f.u.\ below \textit{Oct}. In contrast, after second-stage fine-tuning with PBE+$U$, the recovered model places \textit{Sqr} $0.765$~eV/f.u.\ above \textit{Oct}, reproducing the sign and closely following the profile obtained from direct PBE+$U$ fine-tuning. The inherited source-level phase preference can therefore be reversed, and a compact target-level dataset is sufficient to redirect DPA-4 toward the opposing phase preference. Although both direct and two-stage PBE+$U$ fine-tuning yield a positive \textit{Sqr}--\textit{Oct} energy difference, both underestimate its magnitude, predicting 0.765 eV/f.u.\ compared with the PBE+$U$ value of
0.936 eV/f.u. This residual error is consistent with uneven coverage of the
target dataset. AIMD trajectories initialized near \textit{Sqr} tended to evolve away from this region, while electronic convergence failures further
reduced the number of retained \textit{Sqr}-like configurations. Consequently,
the target dataset contains substantially greater coverage of the \textit{Oct}-like region.
These results therefore demonstrate recovery of the qualitative phase
ordering, whereas quantitatively accurate endpoint energetics may require
targeted enrichment of the underrepresented \textit{Sqr}-like region.

DPA-3 also benefits from second-stage fine-tuning, but its energy and force RMSEs remain substantially higher than those of the DPA-4 models over the investigated training-set sizes, and the greater variation among its independent fine-tuning runs indicates greater sensitivity to the training subset or optimization trajectory under the present fine-tuning protocol. DPA-4 therefore exhibits greater adaptability and reproducibility.

\section{Discussion and Conclusions}

In this work, we investigated whether an MLFF carrying qualitatively incorrect
source-level energetics can be efficiently adapted to a target-level surface. Using NiO as a case study, we considered transfer from a
non-spin-polarized PBE surface, which favors \textit{Sqr} over \textit{Oct}, to an FM PBE+$U$ surface with the opposite \textit{Sqr}--\textit{Oct} ordering. The pretrained DPA-4 models adapted rapidly to the target energies and forces, reaching energy RMSEs below approximately 0.5 meV/atom and force RMSEs near or below 30 meV/$\mathrm{\AA}$ using 170 PBE+$U$ configurations. More importantly, models that had first been fine-tuned to the no-$U$ surface could subsequently recover the opposing PBE+$U$ phase ordering with nearly the same data efficiency as models fine-tuned directly from pretraining. Prior learning of an incorrect phase ordering therefore does not preclude efficient downstream adaptation.

Our results also indicate that downstream adaptability is an important
criterion for evaluating foundation MLFFs. Foundation MLFFs should be assessed not only by their zero-shot errors but also by how efficiently and reproducibly they can be fine-tuned with limited target data. Under the present fine-tuning protocol, the tested DPA-4 models reach lower target errors with fewer configurations and show less variation among independent fine-tuning runs than the tested DPA-3 models. Because the models differ in both architecture and pretraining history, the observed difference cannot be attributed uniquely to architecture. The tested DPA-4 models are more adaptive in the present NiO benchmark.

Several limitations define the scope of these conclusions. First, the
source and target datasets differ in both correlation treatment and magnetic
polarization. The source surface is defined by non-spin-polarized PBE,
whereas the target surface is defined by FM PBE+$U$. The process demonstrated here must therefore be interpreted as transfer between two consistently defined electronic-structure reference surfaces, rather than as an isolated correction from DFT to DFT+$U$. Second, the present results do not establish that arbitrary source-level errors can always be corrected across model architectures, data-selection schemes, and material systems. Transfer may become more difficult when the source data lack relevant coordination environments, the target surface introduces new bonding motifs, or multiple electronic or magnetic states must be represented simultaneously. Finally, an MLFF based only on atomic structure cannot determine the appropriate magnetic state independently. It instead learns the particular electronic-structure branch encoded in its training labels. Consistent magnetic initialization and convergence are therefore essential when constructing target-level datasets for correlated materials.

Together, these observations suggest a division of roles between universal
pretraining and application-specific fine-tuning. Recent work has emphasized
that mixing composition-dependent DFT and DFT+$U$ conventions within a
pretraining dataset can compromise label consistency and has advocated consistent pretraining without U\cite{warfordBetterImpactSelective2026a}. Our results complement this argument by showing that omitting $U$ at the source level need not prevent the resulting model from later adapting to a PBE+$U$ target surface. Pretraining may therefore prioritize broad chemical and configurational coverage, label consistency, and computational affordability, rather than quantitative agreement with the target-level phase ordering of every possible target material. Before system-specific applications, however, a foundation MLFF should be validated against a consistently generated target-level dataset and, when necessary, fine-tuned using a compact set of higher-fidelity calculations. Foundation MLFFs should therefore be viewed not as universally accurate zero-shot calculators but as transferable priors whose practical value depends both on broad prior knowledge and efficient target-level adaptability.

\FloatBarrier
\acknowledgments
This work was supported by the National Key R\&D Program of China (Grant No. 2021YFA0718900), the National Natural Science Foundation of China (Grant Nos. 12574069, 92477114, and 12374096) and the Suzhou Municipal Science and Technology Program (Gusu Innovation and Entrepreneurship Leading Talent Program, Grant No. ZXL2025304). We thank DP Technology for providing computational resources through the Bohrium platform and data hosting services via AIS Square.


\bibliography{refs}

@article{anisimovBandTheoryMott1991,
  title = {Band Theory and {{Mott}} Insulators: {{Hubbard}} {{{\emph{U}}}} Instead of {{Stoner}} {{{\emph{I}}}}},
  shorttitle = {Band Theory and {{Mott}} Insulators},
  author = {Anisimov, Vladimir I. and Zaanen, Jan and Andersen, Ole K.},
  year = 1991,
  month = jul,
  journal = {Physical Review B},
  volume = {44},
  number = {3},
  pages = {943--954},
  issn = {0163-1829, 1095-3795},
  doi = {10.1103/PhysRevB.44.943},
  urldate = {2026-08-09},
  copyright = {http://link.aps.org/licenses/aps-default-license},
  langid = {english}
}

@article{barros-luqueOpenMaterials20242026a,
  title = {The {{Open Materials}} 2024 ({{OMat24}}) Inorganic Materials Dataset and Models},
  author = {{Barros-Luque}, Luis and Shuaibi, Muhammed and Fu, Xiang and Wood, Brandon M. and Dzamba, Misko and Gao, Meng and Rizvi, Ammar and Uyttendaele, Matt and Zitnick, C. Lawrence and Ulissi, Zachary W.},
  year = 2026,
  month = jun,
  journal = {Nature Computational Science},
  volume = {6},
  number = {6},
  pages = {642--652},
  issn = {2662-8457},
  doi = {10.1038/s43588-026-00996-w},
  urldate = {2026-08-09},
  langid = {english}
}

@article{cococcioniLinearResponseApproach2005,
  title = {Linear Response Approach to the Calculation of the Effective Interaction Parameters in the {{LDA}} + {{U}} Method},
  author = {Cococcioni, Matteo and De Gironcoli, Stefano},
  year = 2005,
  month = jan,
  journal = {Physical Review B},
  volume = {71},
  number = {3},
  pages = {035105},
  issn = {1098-0121, 1550-235X},
  doi = {10.1103/PhysRevB.71.035105},
  urldate = {2026-08-09},
  copyright = {http://link.aps.org/licenses/aps-default-license},
  langid = {english}
}

@article{dudarevElectronenergylossSpectraStructural1998a,
  title = {Electron-Energy-Loss Spectra and the Structural Stability of Nickel Oxide: {{An LSDA}}+{{U}} Study},
  shorttitle = {Electron-Energy-Loss Spectra and the Structural Stability of Nickel Oxide},
  author = {Dudarev, S. L. and Botton, G. A. and Savrasov, S. Y. and Humphreys, C. J. and Sutton, A. P.},
  year = 1998,
  month = jan,
  journal = {Physical Review B},
  volume = {57},
  number = {3},
  pages = {1505--1509},
  issn = {0163-1829, 1095-3795},
  doi = {10.1103/PhysRevB.57.1505},
  urldate = {2025-06-05},
  copyright = {http://link.aps.org/licenses/aps-default-license},
  langid = {english}
}

@article{furnessAccurateNumericallyEfficient2020,
  title = {Accurate and {{Numerically Efficient}} R{\textsuperscript{2}} {{SCAN Meta-Generalized Gradient Approximation}}},
  author = {Furness, James W. and Kaplan, Aaron D. and Ning, Jinliang and Perdew, John P. and Sun, Jianwei},
  year = 2020,
  month = oct,
  journal = {The Journal of Physical Chemistry Letters},
  volume = {11},
  number = {19},
  pages = {8208--8215},
  issn = {1948-7185, 1948-7185},
  doi = {10.1021/acs.jpclett.0c02405},
  urldate = {2026-08-09},
  copyright = {https://pubs.acs.org/page/policy/authorchoice\_termsofuse.html},
  langid = {english}
}

@article{hortonAcceleratedDatadrivenMaterials2025,
  title = {Accelerated Data-Driven Materials Science with the {{Materials Project}}},
  author = {Horton, Matthew K. and Huck, Patrick and Yang, Ruo Xi and Munro, Jason M. and Dwaraknath, Shyam and Ganose, Alex M. and Kingsbury, Ryan S. and Wen, Mingjian and Shen, Jimmy X. and Mathis, Tyler S. and Kaplan, Aaron D. and Berket, Karlo and Riebesell, Janosh and George, Janine and Rosen, Andrew S. and {Spotte-Smith}, Evan W. C. and McDermott, Matthew J. and Cohen, Orion A. and Dunn, Alex and Kuner, Matthew C. and Rignanese, Gian-Marco and Petretto, Guido and Waroquiers, David and Griffin, Sinead M. and Neaton, Jeffrey B. and Chrzan, Daryl C. and Asta, Mark and Hautier, Geoffroy and Cholia, Shreyas and Ceder, Gerbrand and Ong, Shyue Ping and Jain, Anubhav and Persson, Kristin A.},
  year = 2025,
  month = jul,
  journal = {Nature Materials},
  issn = {1476-1122, 1476-4660},
  doi = {10.1038/s41563-025-02272-0},
  urldate = {2025-08-14},
  langid = {english}
}

@article{jainCommentaryMaterialsProject2013,
  title = {Commentary: {{The Materials Project}}: {{A}} Materials Genome Approach to Accelerating Materials Innovation},
  shorttitle = {Commentary},
  author = {Jain, Anubhav and Ong, Shyue Ping and Hautier, Geoffroy and Chen, Wei and Richards, William Davidson and Dacek, Stephen and Cholia, Shreyas and Gunter, Dan and Skinner, David and Ceder, Gerbrand and Persson, Kristin A.},
  year = 2013,
  month = jul,
  journal = {APL Materials},
  volume = {1},
  number = {1},
  pages = {011002},
  issn = {2166-532X},
  doi = {10.1063/1.4812323},
  urldate = {2024-10-15},
  langid = {english}
}

@article{jainFormationEnthalpiesMixing2011,
  title = {Formation Enthalpies by Mixing {{GGA}} and {{GGA}} + {{U}} Calculations},
  author = {Jain, Anubhav and Hautier, Geoffroy and Ong, Shyue Ping and Moore, Charles J. and Fischer, Christopher C. and Persson, Kristin A. and Ceder, Gerbrand},
  year = 2011,
  month = jul,
  journal = {Physical Review B},
  volume = {84},
  number = {4},
  pages = {045115},
  issn = {1098-0121, 1550-235X},
  doi = {10.1103/PhysRevB.84.045115},
  urldate = {2026-08-09},
  copyright = {http://link.aps.org/licenses/aps-default-license},
  langid = {english}
}

@article{kamInterplayElectronLocalization2025,
  title = {Interplay between Electron Localization, Magnetic Order, and {{Jahn-Teller}} Distortion Dictates {{LiMnO}} 2 Phase Stability},
  author = {Kam, Ronald L. and Binci, Luca and Kaplan, Aaron D. and Persson, Kristin A. and Marzari, Nicola and Ceder, Gerbrand},
  year = 2025,
  month = jun,
  journal = {Physical Review B},
  volume = {111},
  number = {24},
  pages = {245132},
  issn = {2469-9950, 2469-9969},
  doi = {10.1103/99jn-17v6},
  urldate = {2025-06-20},
  langid = {english}
}

@misc{kaplanFoundationalPotentialEnergy2025,
  title = {A {{Foundational Potential Energy Surface Dataset}} for {{Materials}}},
  author = {Kaplan, Aaron D. and Liu, Runze and Qi, Ji and Ko, Tsz Wai and Deng, Bowen and Riebesell, Janosh and Ceder, Gerbrand and Persson, Kristin A. and Ong, Shyue Ping},
  year = 2025,
  month = mar,
  number = {arXiv:2503.04070},
  eprint = {2503.04070},
  primaryclass = {cond-mat},
  publisher = {arXiv},
  doi = {10.48550/arXiv.2503.04070},
  urldate = {2025-03-10},
  archiveprefix = {arXiv},
  langid = {english}
}

@article{kresseEfficiencyAbinitioTotal1996,
  title = {Efficiency of Ab-Initio Total Energy Calculations for Metals and Semiconductors Using a Plane-Wave Basis Set},
  author = {Kresse, G. and Furthm{\"u}ller, J.},
  year = 1996,
  month = jul,
  journal = {Computational Materials Science},
  volume = {6},
  number = {1},
  pages = {15--50},
  issn = {09270256},
  doi = {10.1016/0927-0256(96)00008-0},
  urldate = {2025-06-05},
  copyright = {https://www.elsevier.com/tdm/userlicense/1.0/},
  langid = {english}
}

@article{kresseEfficientIterativeSchemes1996,
  title = {Efficient Iterative Schemes for {\emph{Ab Initio}} Total-Energy Calculations Using a Plane-Wave Basis Set},
  author = {Kresse, G. and Furthm{\"u}ller, J.},
  year = 1996,
  month = oct,
  journal = {Physical Review B},
  volume = {54},
  number = {16},
  pages = {11169--11186},
  issn = {0163-1829, 1095-3795},
  doi = {10.1103/PhysRevB.54.11169},
  urldate = {2025-06-05},
  copyright = {http://link.aps.org/licenses/aps-default-license},
  langid = {english}
}

@article{kresseInitioMolecularDynamics1993,
  title = {Ab Initio Molecular Dynamics for Liquid Metals},
  author = {Kresse, G. and Hafner, J.},
  year = 1993,
  month = jan,
  journal = {Physical Review B},
  volume = {47},
  number = {1},
  pages = {558--561},
  publisher = {American Physical Society},
  doi = {10.1103/PhysRevB.47.558},
  urldate = {2025-06-05}
}

@article{kresseUltrasoftPseudopotentialsProjector1999,
  title = {From Ultrasoft Pseudopotentials to the Projector Augmented-Wave Method},
  author = {Kresse, G. and Joubert, D.},
  year = 1999,
  month = jan,
  journal = {Physical Review B},
  volume = {59},
  number = {3},
  pages = {1758--1775},
  issn = {0163-1829, 1095-3795},
  doi = {10.1103/PhysRevB.59.1758},
  urldate = {2025-06-05},
  copyright = {http://link.aps.org/licenses/aps-default-license},
  langid = {english}
}

@article{krukauInfluenceExchangeScreening2006,
  title = {Influence of the Exchange Screening Parameter on the Performance of Screened Hybrid Functionals},
  author = {Krukau, Aliaksandr V. and Vydrov, Oleg A. and Izmaylov, Artur F. and Scuseria, Gustavo E.},
  year = 2006,
  month = dec,
  journal = {The Journal of Chemical Physics},
  volume = {125},
  number = {22},
  pages = {224106},
  issn = {0021-9606, 1089-7690},
  doi = {10.1063/1.2404663},
  urldate = {2026-08-09},
  langid = {english}
}

@article{kunerMPALOER2SCANDataset2025a,
  title = {{{MP-ALOE}}: An {{r2SCAN}} Dataset for Universal Machine Learning Interatomic Potentials},
  shorttitle = {{{MP-ALOE}}},
  author = {Kuner, Matthew C. and Kaplan, Aaron D. and Persson, Kristin A. and Asta, Mark and Chrzan, Daryl C.},
  year = 2025,
  month = nov,
  journal = {npj Computational Materials},
  volume = {11},
  number = {1},
  pages = {352},
  issn = {2057-3960},
  doi = {10.1038/s41524-025-01834-9},
  urldate = {2026-08-09},
  langid = {english}
}

@misc{liDPA4PushingAccuracyCost2026,
  title = {{{DPA4}}: {{Pushing}} the {{Accuracy-Cost Frontier}} of {{Interatomic Potentials}} with {{EMFA SO}}(2) {{Convolution}}},
  shorttitle = {{{DPA4}}},
  author = {Li, Tiancheng and Li, Wentao and Peng, Anyang and Xue, Jianming and Zhang, Linfeng and Zhang, Duo and Wang, Han},
  year = 2026,
  month = jun,
  number = {arXiv:2606.02419},
  eprint = {2606.02419},
  primaryclass = {physics.chem-ph},
  publisher = {arXiv},
  doi = {10.48550/arXiv.2606.02419},
  urldate = {2026-06-02},
  archiveprefix = {arXiv},
  langid = {english}
}

@article{longEvaluatingOptimal32020,
  title = {Evaluating Optimal {{U}} for 3 d Transition-Metal Oxides within the {{SCAN}}+ {{U}} Framework},
  author = {Long, Olivia Y. and Sai Gautam, Gopalakrishnan and Carter, Emily A.},
  year = 2020,
  month = apr,
  journal = {Physical Review Materials},
  volume = {4},
  number = {4},
  pages = {045401},
  issn = {2475-9953},
  doi = {10.1103/PhysRevMaterials.4.045401},
  urldate = {2026-08-09},
  langid = {english}
}

@misc{Models343DPA313M,
  title        = {{DPA-3.1-3M model}},
  howpublished = {AIS Square model repository},
  url          = {https://www.aissquare.com/models/detail?pageType=models&name=DPA-3.1-3M&id=343},
  urldate      = {2025-10-19}
}

@article{ongPythonMaterialsGenomics2013,
  title = {Python {{Materials Genomics}} (Pymatgen): {{A}} Robust, Open-Source Python Library for Materials Analysis},
  shorttitle = {Python {{Materials Genomics}} (Pymatgen)},
  author = {Ong, Shyue Ping and Richards, William Davidson and Jain, Anubhav and Hautier, Geoffroy and Kocher, Michael and Cholia, Shreyas and Gunter, Dan and Chevrier, Vincent L. and Persson, Kristin A. and Ceder, Gerbrand},
  year = 2013,
  month = feb,
  journal = {Computational Materials Science},
  volume = {68},
  pages = {314--319},
  issn = {09270256},
  doi = {10.1016/j.commatsci.2012.10.028},
  urldate = {2024-10-15},
  copyright = {https://www.elsevier.com/tdm/userlicense/1.0/},
  langid = {english}
}

@article{perdewGeneralizedGradientApproximation1996,
  title = {Generalized {{Gradient Approximation Made Simple}}},
  author = {Perdew, John P. and Burke, Kieron and Ernzerhof, Matthias},
  year = 1996,
  month = oct,
  journal = {Physical Review Letters},
  volume = {77},
  number = {18},
  pages = {3865--3868},
  issn = {0031-9007, 1079-7114},
  doi = {10.1103/PhysRevLett.77.3865},
  urldate = {2025-06-05},
  copyright = {http://link.aps.org/licenses/aps-default-license},
  langid = {english}
}

@article{perdewGeneralizedGradientApproximation1997,
  title = {Generalized {{Gradient Approximation Made Simple}} [{{Phys}}. {{Rev}}. {{Lett}}. 77, 3865 (1996)]},
  author = {Perdew, John P. and Burke, Kieron and Ernzerhof, Matthias},
  year = 1997,
  month = feb,
  journal = {Physical Review Letters},
  volume = {78},
  number = {7},
  pages = {1396--1396},
  issn = {0031-9007, 1079-7114},
  doi = {10.1103/PhysRevLett.78.1396},
  urldate = {2025-06-05},
  copyright = {http://link.aps.org/licenses/aps-default-license},
  langid = {english}
}

@misc{QUESTS20240621Documentation,
  title = {{{QUESTS}} 2024.06.21 Documentation},
  urldate = {2026-08-09},
  url = {https://dskoda.github.io/quests/},
  howpublished = {github.io}
}

@article{schmidtDataset175kStable2022,
  title = {A Dataset of 175k Stable and Metastable Materials Calculated with the {{PBEsol}} and {{SCAN}} Functionals},
  author = {Schmidt, Jonathan and Wang, Hai-Chen and Cerqueira, Tiago F. T. and Botti, Silvana and Marques, Miguel A. L.},
  year = 2022,
  month = mar,
  journal = {Scientific Data},
  volume = {9},
  number = {1},
  pages = {64},
  publisher = {Nature Publishing Group},
  issn = {2052-4463},
  doi = {10.1038/s41597-022-01177-w},
  urldate = {2025-10-18},
  copyright = {2022 The Author(s)},
  langid = {english}
}

@article{schmidtMachineLearningAssistedDeterminationGlobal2023,
  title = {Machine-{{Learning-Assisted Determination}} of the {{Global Zero-Temperature Phase Diagram}} of {{Materials}}},
  author = {Schmidt, Jonathan and Hoffmann, Noah and Wang, Hai-Chen and Borlido, Pedro and Carri{\c c}o, Pedro J. M. A. and Cerqueira, Tiago F. T. and Botti, Silvana and Marques, Miguel A. L.},
  year = 2023,
  journal = {Advanced Materials},
  volume = {35},
  number = {22},
  pages = {2210788},
  issn = {1521-4095},
  doi = {10.1002/adma.202210788},
  urldate = {2025-10-18},
  langid = {english}
}

@article{schwalbe-kodaModelfreeEstimationCompleteness2025,
  title = {Model-Free Estimation of Completeness, Uncertainties, and Outliers in Atomistic Machine Learning Using Information Theory},
  author = {{Schwalbe-Koda}, Daniel and Hamel, Sebastien and Sadigh, Babak and Zhou, Fei and Lordi, Vincenzo},
  year = 2025,
  month = apr,
  journal = {Nature Communications},
  volume = {16},
  number = {1},
  pages = {4014},
  issn = {2041-1723},
  doi = {10.1038/s41467-025-59232-0},
  urldate = {2025-05-03},
  langid = {english}
}

@misc{shinagawaMatlantisPFPV8Universal2026,
  title = {Matlantis-{{PFP}} v8: {{Universal Machine Learning Interatomic Potential}} with {{Better Experimental Agreements}} via {{r2SCAN Functional}}},
  shorttitle = {Matlantis-{{PFP}} V8},
  author = {Shinagawa, Chikashi and Takamoto, So and Shintani, Daiki and Zhuang, Yong-Bin and Tsuboi, Yuta and Nishimra, Katsuhiko and Shinohara, Kohei and Iwase, Shigeru and Tanaka, Yuta and Li, Ju},
  year = 2026,
  month = mar,
  number = {arXiv:2603.11063},
  eprint = {2603.11063},
  primaryclass = {physics.chem-ph},
  publisher = {arXiv},
  doi = {10.48550/arXiv.2603.11063},
  urldate = {2026-08-09},
  archiveprefix = {arXiv}
}

@article{spagnoliDensityFunctionalTheory2010,
  title = {Density Functional Theory Study of the Relative Stability of the Iron Disulfide Polymorphs Pyrite and Marcasite},
  author = {Spagnoli, D. and Refson, K. and Wright, K. and Gale, J. D.},
  year = 2010,
  month = mar,
  journal = {Physical Review B},
  volume = {81},
  number = {9},
  pages = {094106},
  issn = {1098-0121, 1550-235X},
  doi = {10.1103/PhysRevB.81.094106},
  urldate = {2026-08-09},
  copyright = {http://link.aps.org/licenses/aps-default-license},
  langid = {english}
}

@article{wangPretrainingFinetuningDistillation2025,
  title = {Pre-Training, Fine-Tuning, and Distillation ({{PFD}}): {{Automatically}} Generating Machine Learning Force Fields from Universal Models},
  shorttitle = {Pre-Training, Fine-Tuning, and Distillation ({{PFD}})},
  author = {Wang, Ruoyu and Gao, Yuxiang and Wu, Hongyu and Zhong, Zhicheng},
  year = 2025,
  month = nov,
  journal = {Physical Review Materials},
  volume = {9},
  number = {11},
  pages = {113802},
  issn = {2475-9953},
  doi = {10.1103/sbz6-btz8},
  urldate = {2026-02-11},
  langid = {english}
}

@article{warfordBetterImpactSelective2026a,
  title = {Better without {{U}} : Impact of Selective {{Hubbard U}} Correction on Foundational {{MLIPs}}},
  shorttitle = {Better without {{U}}},
  author = {Warford, Thomas and Thiemann, Fabian L and Cs{\'a}nyi, G{\'a}bor},
  year = 2026,
  month = jun,
  journal = {Machine Learning: Science and Technology},
  volume = {7},
  number = {3},
  pages = {035033},
  issn = {2632-2153},
  doi = {10.1088/2632-2153/ae6be5},
  urldate = {2026-08-09},
  langid = {english}
}

@misc{zengDeePMDkitV3MultipleBackend2025,
  title = {{{DeePMD-kit}} v3: {{A Multiple-Backend Framework}} for {{Machine Learning Potentials}}},
  shorttitle = {{{DeePMD-kit}} V3},
  author = {Zeng, Jinzhe and Zhang, Duo and Peng, Anyang and Zhang, Xiangyu and He, Sensen and Wang, Yan and Liu, Xinzijian and Bi, Hangrui and Li, Yifan and Cai, Chun and Zhang, Chengqian and Du, Yiming and Zhu, Jia-Xin and Mo, Pinghui and Huang, Zhengtao and Zeng, Qiyu and Shi, Shaochen and Qin, Xuejian and Yu, Zhaoxi and Luo, Chenxing and Ding, Ye and Liu, Yun-Pei and Shi, Ruosong and Wang, Zhenyu and Bore, Sigbj{\o}rn L{\o}land and Chang, Junhan and Deng, Zhe and Ding, Zhaohan and Han, Siyuan and Jiang, Wanrun and Ke, Guolin and Liu, Zhaoqing and Lu, Denghui and Muraoka, Koki and Oliaei, Hananeh and Singh, Anurag Kumar and Que, Haohui and Xu, Weihong and Xu, Zhangmancang and Zhuang, Yong-Bin and Dai, Jiayu and Giese, Timothy J. and Jia, Weile and Xu, Ben and York, Darrin M. and Zhang, Linfeng and Wang, Han},
  year = 2025,
  month = feb,
  number = {arXiv:2502.19161},
  eprint = {2502.19161},
  primaryclass = {physics},
  publisher = {arXiv},
  doi = {10.48550/arXiv.2502.19161},
  urldate = {2025-03-03},
  archiveprefix = {arXiv},
  langid = {english}
}

@article{zhangGraphNeuralNetwork2026,
  title = {A Graph Neural Network for the Era of Large Atomistic Models},
  author = {Zhang, Duo and Peng, Anyang and Cai, Chun and Li, Wentao and Zhou, Yuanchang and Zeng, Jinzhe and Guo, Mingyu and Zhang, Chengqian and Li, Bowen and Jiang, Hong and Zhu, Tong and Jia, Weile and Zhang, Linfeng and Wang, Han},
  year = 2026,
  month = may,
  journal = {npj Computational Materials},
  issn = {2057-3960},
  doi = {10.1038/s41524-026-02146-2},
  urldate = {2026-08-09},
  langid = {english}
}

@article{zhouFirstprinciplesPredictionRedox2004,
  title = {First-Principles Prediction of Redox Potentials in Transition-Metal Compounds with {{LDA}} + {{U}}},
  author = {Zhou, F. and Cococcioni, M. and Marianetti, C. A. and Morgan, D. and Ceder, G.},
  year = 2004,
  month = dec,
  journal = {Physical Review B},
  volume = {70},
  number = {23},
  pages = {235121},
  issn = {1098-0121, 1550-235X},
  doi = {10.1103/PhysRevB.70.235121},
  urldate = {2026-08-09},
  copyright = {http://link.aps.org/licenses/aps-default-license},
  langid = {english}
}

\end{document}